\documentclass[cameraready]{Interspeech}

\usepackage{cite}
\usepackage{flushend}
\usepackage{tabularx}  
\title{An Objective Intelligibility Metric Evaluation on Spanish Speech}

\author[affiliation={1}, orcid=0000-0001-8634-7897]{Iván}{López-Espejo}
\author[affiliation={2,3}, orcid=0000-0003-1478-622X]{Jesper}{Jensen}

\address{
    $^1$ Department of Signal Theory, Telematics and Communications, University of Granada, Spain \\
    $^2$ Department of Electronic Systems, Aalborg University, Denmark \\
    $^3$ Oticon A/S, Denmark
}

\email{iloes@ugr.es, jesj@demant.com}


\usepackage{comment}

\begin{document}

\maketitle

\begin{abstract}
	Objective intelligibility metrics (OIMs) enable fast and low-cost evaluation of speech intelligibility and are widely used in speech technology assessment. In this study, we evaluate five reference-based OIMs (STOI, ESTOI, STGI, HASPI, and SIIB) and two deep learning-based no-reference metrics (MOSA-Net+ and W2V-SIP) on SpInt, a new Spanish speech intelligibility dataset. Our results show that reference-based OIMs consistently outperform modern data-driven no-reference approaches, which degrade notably under training--test acoustic mismatches such as language mismatch. This effect is particularly relevant in our scenario, as none of the evaluated metrics were exposed to Spanish speech data during development. Consequently, to foster research on more robust and generalizable no-reference OIMs, SpInt is released publicly.
\end{abstract}

\section{Introduction}
\label{sec:intro}

Speech intelligibility---defined as the listener's ability to accurately perceive and understand spoken language---is a critical factor in the development and evaluation of speech-related technologies, including speech enhancement systems, hearing aids, and speech synthesis \cite{Haolan26,Yang25}. Speech intelligibility is commonly quantified as the proportion of correctly recognized linguistic units (e.g., words \cite{Allen}) and is ideally assessed through listening tests with panels of participants; however, such evaluations are typically time-consuming, labor-intensive, and costly. This has motivated the development of objective metrics that can reliably predict human speech intelligibility as an alternative to listening tests and constitutes an important area of ongoing research \cite{Haolan26}.

Objective intelligibility metrics (OIMs) can be broadly categorized into reference-based (intrusive) and no-reference (non-intrusive) approaches. Reference-based OIMs estimate the intelligibility of a degraded (noisy, distorted, and/or processed) signal by comparing it to a clean reference speech signal. In contrast, no-reference OIMs predict speech intelligibility from the degraded signal alone, without access to a clean reference. Recent advances in deep learning have accelerated the development of no-reference OIMs, enabling promising performance by learning intelligibility-related features from data \cite{Pedersen23,MOSA-Net+,Haolan,Haolan26}. Developing accurate no-reference OIMs is therefore crucial to expanding their applicability in real-world scenarios, where clean reference signals are typically unavailable.

In this study, we evaluate the performance of five reference-based and two no-reference OIMs on a novel Spanish speech intelligibility dataset, referred to as SpInt. Unlike prior work that conducts comparative evaluations of OIMs (e.g., \cite{VanKuyk}), to the best of our knowledge, this is the first study to assess both intrusive and modern deep learning-based non-intrusive OIMs for the Spanish language.

While modern data-driven no-reference OIMs have surpassed reference-based OIMs on some speech intelligibility datasets, their performance degrades significantly under mismatched conditions (e.g., language or noise type) between training and test sets \cite{Pedersen23,Haolan,Haolan26}. Since all seven OIMs evaluated in this study were developed using speech data in languages other than Spanish, our main finding is therefore unsurprising: intrusive OIMs tend to perform better in our experimental framework, likely due to their exploitation of information from clean reference signals. Hence, to support the development of more robust and generalizable no-reference OIMs, we make SpInt publicly available\footnote{\url{https://doi.org/10.5281/zenodo.20763579}\label{spint}}, thereby helping to alleviate the data scarcity that currently limits progress in this area \cite{Pedersen23}.

The remainder of this paper is organized as follows. The speech intelligibility dataset SpInt is described in Section \ref{sec:spint}. Section \ref{sec:results} presents and evaluates the OIMs, and discusses the results. Finally, Section \ref{sec:conclusion} concludes the paper.

\section{SpInt: A Spanish speech intelligibility dataset}
\label{sec:spint}

SpInt is a Spanish speech intelligibility dataset collected using a standard closed-vocabulary Hagerman test \cite{Hagerman82,Hagerman}, a matrix sentence test comprising five-word sentences with a fixed syntactic structure: name--verb--number--noun--adjective. Each sentence is generated by randomly sampling from a set of 50 words (10 per syntactic category), resulting in $10^5$ possible grammatically valid combinations.

SpInt contains 5,148 utterances paired with behavioral intelligibility scores and is publicly available\footref{spint}. The following subsections describe the characteristics of these utterances and the procedure used to obtain the corresponding scores, which closely follow the methodology in \cite{Espejo26}.

\subsection{Listening test stimuli}
\label{ssec:stimuli}

The intelligibility test stimuli were generated using the Spanish version of the Oldenburg Sentence Test (OLSA) speech material \cite{Sabine12}. The dataset consists of 100 utterances recorded in a quiet environment by a single female native speaker of Spanish.

Clean speech signals were downsampled to 16 kHz and additively corrupted with bus station noise recorded in Granada (Spain) \cite{Juanma20} at multiple signal-to-noise ratios (SNRs). Eleven SNR conditions were considered: \{-15, -12.5, -10, -7.5, -5, -2.5, 0, 2.5, 5, 10, $+\infty$\} dB.

In addition to the noisy (unprocessed) condition, stimuli processed by two state-of-the-art deep learning-based speech enhancement systems were also included: FullSubNet+ \cite{FullSubNet+}, a complex spectral masking-based discriminative method\footnote{\url{https://github.com/RookieJunChen/FullSubNet-plus}}, and SGMSE+ \cite{SGMSE+}, a diffusion-based generative model\footnote{\url{https://github.com/sp-uhh/sgmse}}, for both of which official pre-trained checkpoints were employed.

For each combination of SNR and processing condition (i.e., unprocessed, FullSubNet+, or SGMSE+), six different sentences were used, resulting in a total of 1 noise $\times$ 11 SNRs $\times$ 3 processing conditions $\times$ 6 sentences = 198 stimuli. The exact realization of the stimuli was randomized across participants.

\begin{figure}[t]
	\centering
	\includegraphics[width=0.95\linewidth]{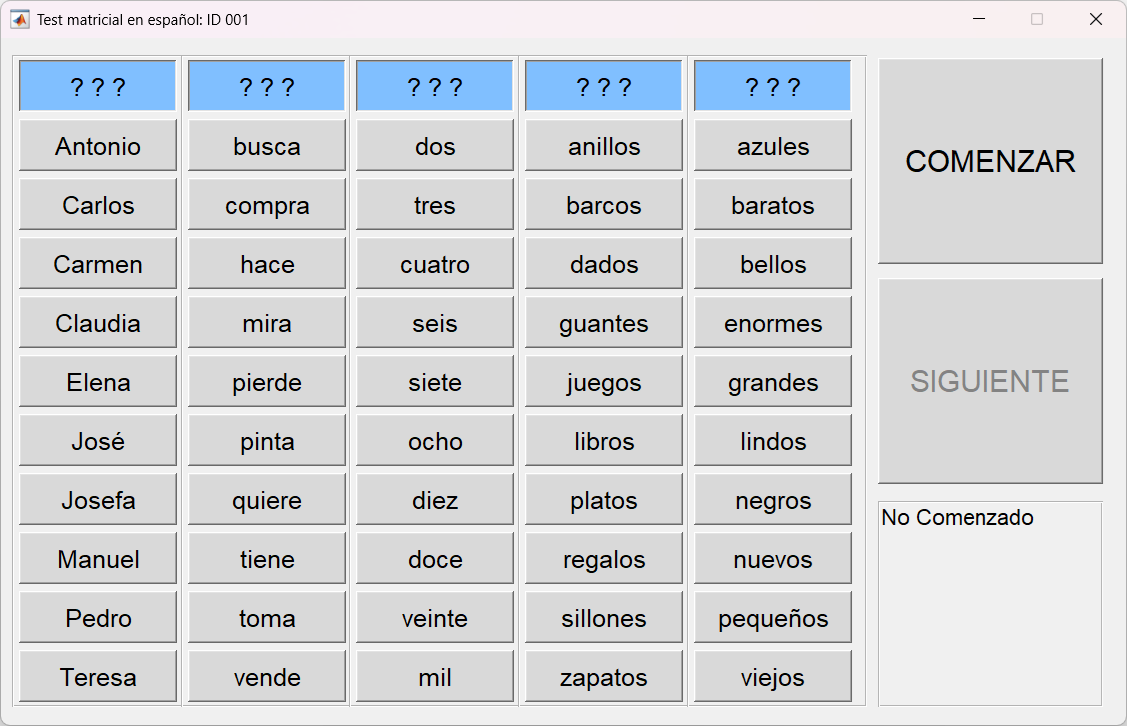}
	\caption{Graphical user interface of the listening test, showing the 10$\times$5 word matrix used for response selection.}
	\label{fig:test}
\end{figure}

\subsection{Listening test procedure}
\label{ssec:procedure}

The listening experiment was conducted in a quiet room at CETIC-UGR (Granada, Spain). Participants were seated at a desk and listened to the stimuli through Beyerdynamic DT 770 Pro headphones driven by a Focusrite Scarlett 2i2 audio interface, connected to an HP ProBook 450 15.6-inch G10 laptop displaying the test interface. The interface presented all candidate words for each of the five syntactic categories in a 10$\times$5 matrix (see Figure \ref{fig:test}). Each participant initiated stimulus playback---presented only once---via a mouse click and then selected the perceived words from the matrix. After completing their selection, the participant proceeded to the next stimulus by clicking a button. This procedure was repeated for multiple stimuli until the end of the test.

Twenty-six native speakers of Spanish (13 female, 13 male) with self-reported normal hearing participated in the study. While the majority were from Spain, several participants were from Spanish-speaking countries in Latin America. Participants ranged in age from 19 to 63 years (mean = 33.6 years, SD = 13.5 years).

Prior to the test, participants were instructed to identify each word as accurately as possible and to choose the default ``\emph{do not know}'' option only when they would otherwise be guessing. The presentation level was individually adjusted to a comfortable listening level at the beginning of the session and kept constant thereafter. The experiment began with a training phase consisting of 18 stimuli (generated using a procedure analogous to that described in Subsection \ref{ssec:stimuli}), allowing participants to become familiar with the interface and reducing potential learning effects \cite{Tytti23}. The main test comprised 198 stimuli, divided into two equal blocks of 99 items separated by a short break to mitigate listening fatigue. Stimulus order was randomized across participants. On average, the training phase lasted 3--4 minutes, while the first and second test blocks required approximately 18--19 and 17--18 minutes, respectively. Participants received financial compensation for their time.

\subsection{Behavioral intelligibility results}

\begin{figure}[t]
	\centering
	\includegraphics[width=0.9\linewidth]{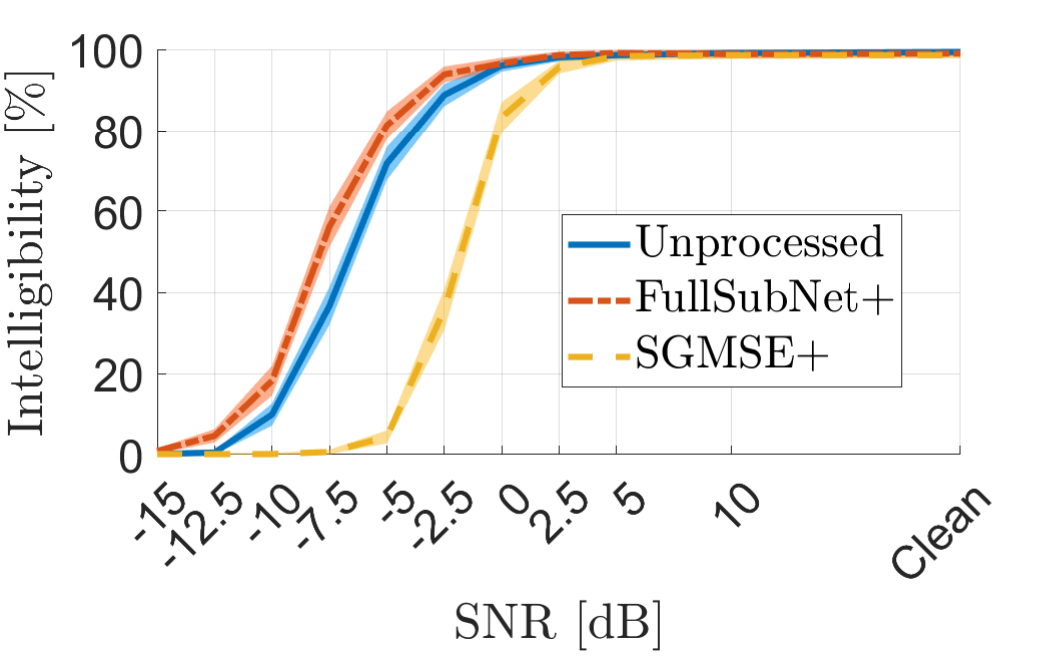}
	\caption{Behavioral intelligibility (\%) as a function of SNR and processing condition. Shaded areas indicate 95\% confidence intervals based on the Student's $t$-distribution \cite{CasellaBerger2002}, computed by pooling all sentence-level trials across participants within each SNR and processing-condition combination ($N=156$ trials per combination).}
	\label{fig:int}
\end{figure}

Figure \ref{fig:int} shows mean behavioral intelligibility---computed as the percentage of correctly recognized words \cite{Allen}---from the listening test described in Subsection \ref{ssec:procedure}, aggregated across participants for each SNR and processing condition. As expected, intelligibility increases monotonically with SNR. FullSubNet+ yields a modest improvement over the unprocessed noisy signals, whereas SGMSE+ substantially degrades intelligibility at low SNRs. This degradation may stem from several factors, including language mismatch (SGMSE+ was trained on English data only), a limited training SNR range ($[0, 20]$ dB), and the tendency of diffusion-based models to hallucinate at very low SNRs \cite{Genhancer,SGMSE+}. It is important to emphasize that this substantial degradation is not reflected in OIMs such as ESTOI \cite{ESTOI}, which in this work (results not shown for clarity) and in previous studies \cite{Espejo26} generally predict improved intelligibility for SGMSE+ processed signals, with the exception of extremely low SNRs.

\section{Evaluation of objective intelligibility metrics on SpInt}
\label{sec:results}

\begin{figure*}[t]
	\centering
	\includegraphics[height=3.1cm]{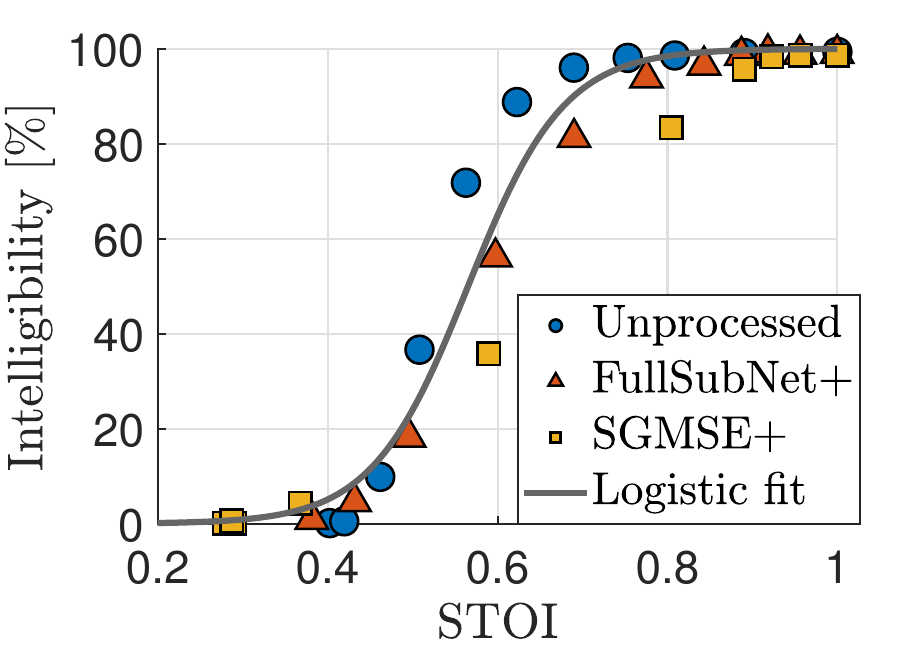}
	\includegraphics[height=3.1cm]{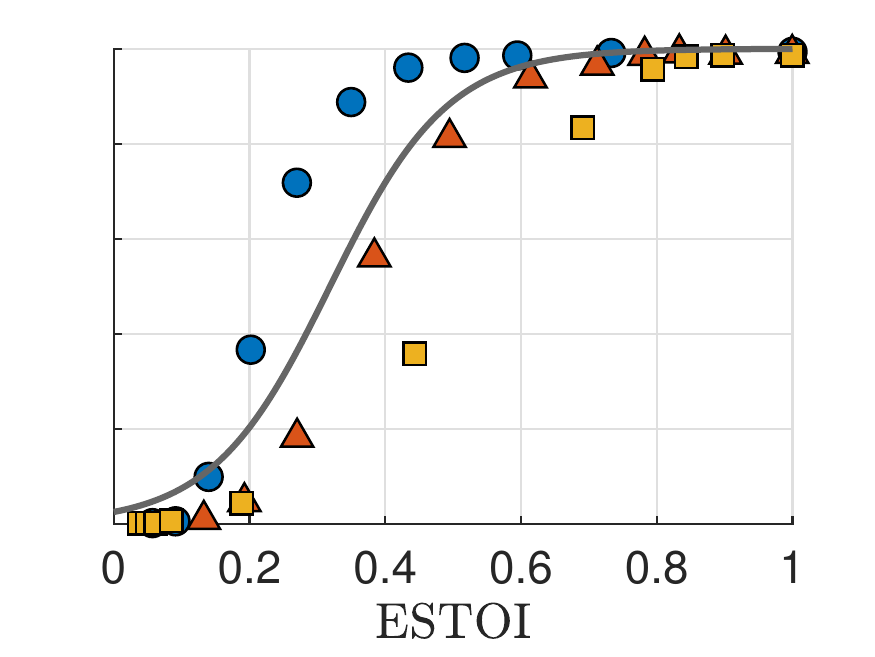}
	\includegraphics[height=3.1cm]{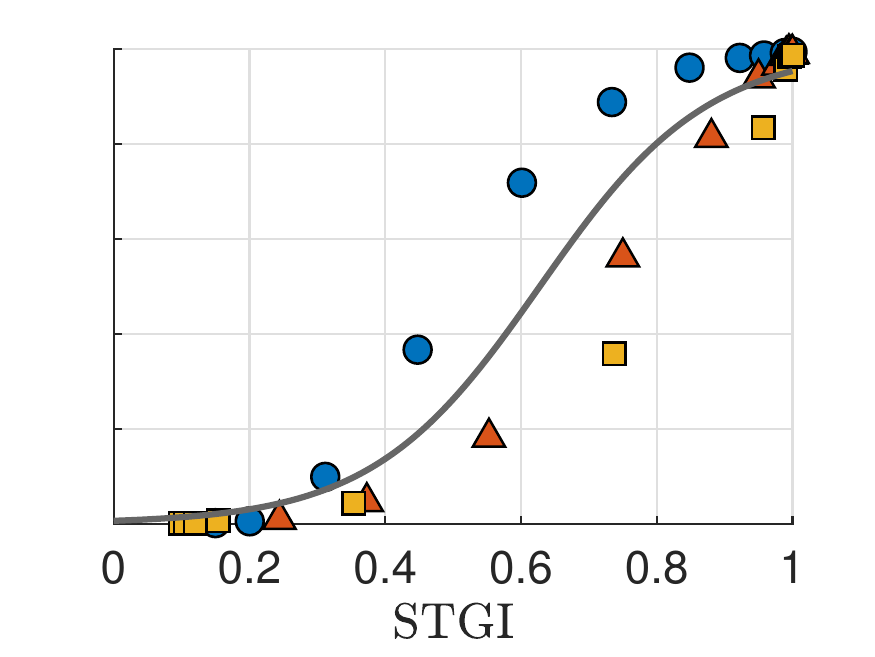}
	\includegraphics[height=3.1cm]{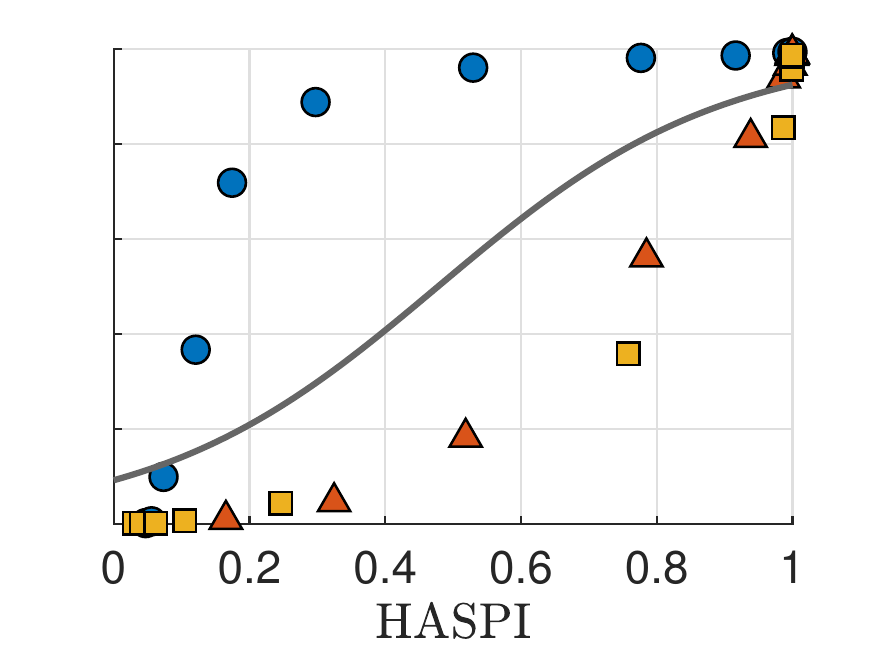} \\
	\includegraphics[height=3.1cm]{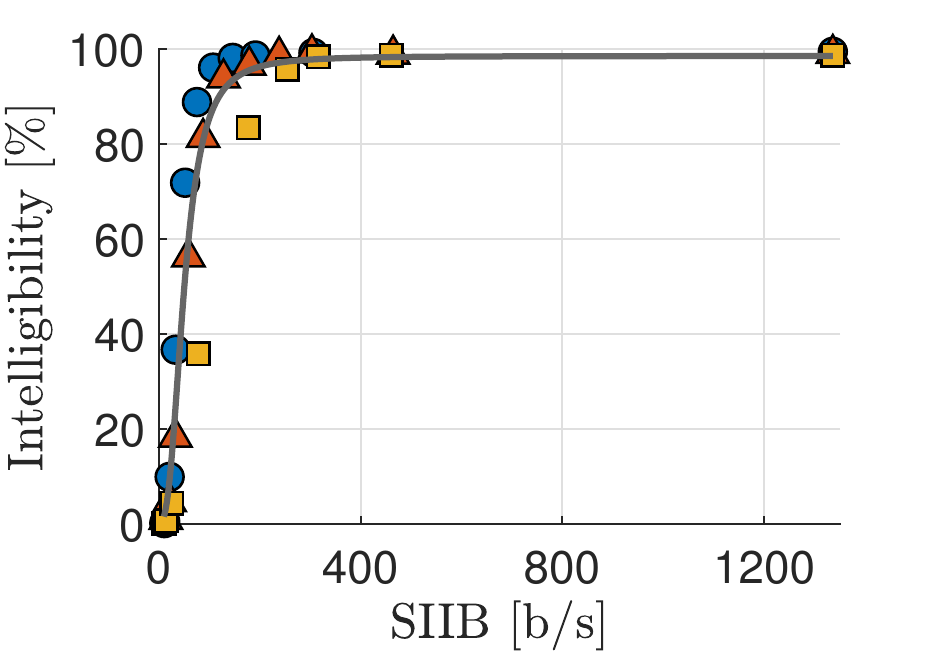}
	\includegraphics[height=3.1cm]{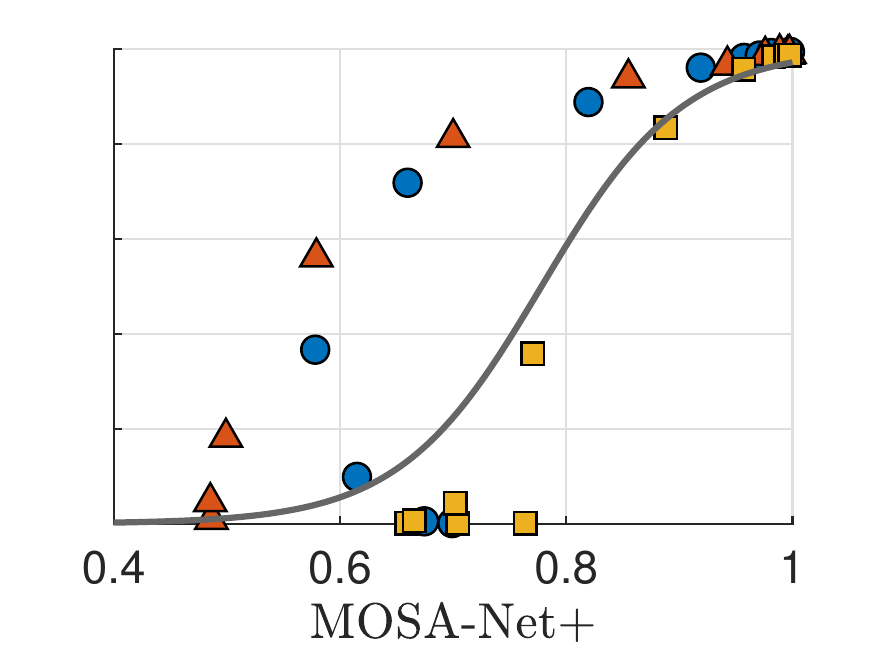}
	\includegraphics[height=3.1cm]{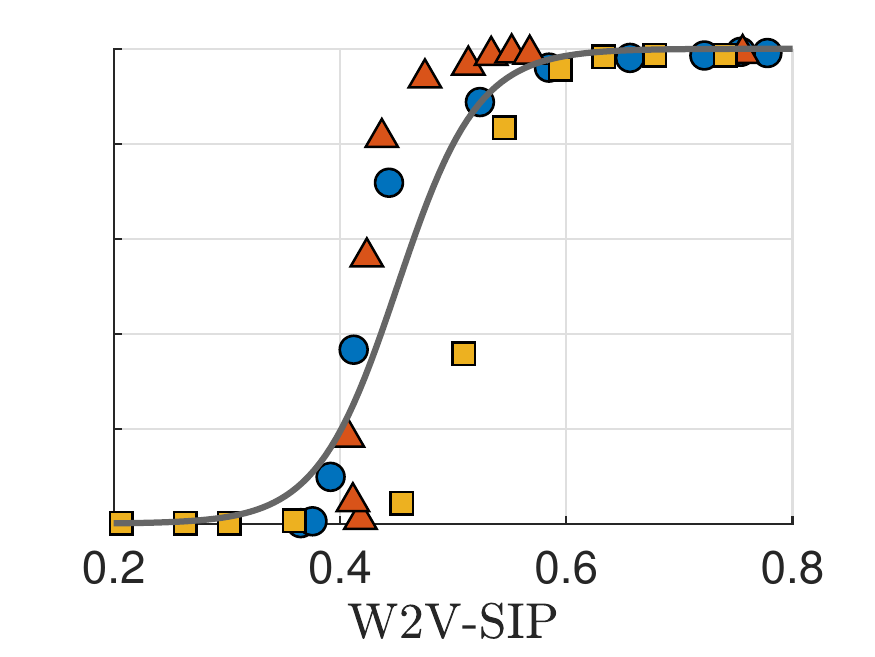}
	\caption{Scatter plots of behavioral intelligibility (\%) versus the scores from the evaluated OIMs. Each data point corresponds to a combination of SNR and processing condition. Logistic fits are shown for each metric.}
	\label{fig:scatter}
\end{figure*}

In this section, we evaluate seven OIMs on SpInt. Subsection \ref{ssec:oims} presents the considered metrics, while Subsection \ref{ssec:results} reports and discusses the results.

\subsection{Objective intelligibility metrics}
\label{ssec:oims}

In this work, we assess the following five reference-based OIMs:
\begin{enumerate}
	\item Short-time objective intelligibility (\textbf{STOI}) \cite{STOI} estimates speech intelligibility by measuring the short-time correlation between temporal envelopes of clean and degraded speech in frequency bands.
	\item Extended STOI (\textbf{ESTOI}) \cite{ESTOI} is an extension of STOI that computes correlations between short-time spectral representations of clean and degraded speech, improving robustness to temporally modulated noise.
	\item Spectro-temporal glimpsing index (\textbf{STGI}) \cite{STGI} is based on detecting glimpses in short-time segments obtained from a spectro-temporal modulation decomposition of the input speech signals.
	\item Hearing-aid speech perception index (\textbf{HASPI}) \cite{HASPI} is based on an auditory model in which the shape and bandwidth of cochlear filters depend on speech signal intensity and the degree of outer hair-cell impairment of the listener. For a fair comparison, HASPI is applied in normal-hearing listener mode in this work.
	\item Speech intelligibility in bits (\textbf{SIIB}) \cite{SIIB} is an information-theoretic OIM that uses a non-parametric mutual information estimator to quantify the information shared between clean and degraded speech signals.
\end{enumerate}

In addition, the following two modern deep learning-based no-reference OIMs are also evaluated:
\begin{enumerate}
	\item The multi-objective speech assessment model (\textbf{MOSA-Net+}) \cite{MOSA-Net+} predicts subjective quality and intelligibility scores in a multi-task framework by combining features from Whisper \cite{Whisper} with traditional spectral features and learnable filterbank representations.
	\item The speech intelligibility predictor in \cite{Haolan} (\textbf{W2V-SIP}) exploits features from a wav2vec 2.0 \cite{W2V2} model adapted for noise-robust automatic speech recognition (ASR) and exhibits strong generalization to unseen noise conditions.
\end{enumerate}
It is worth noting that these two non-intrusive OIMs use ASR models to extract features for speech intelligibility prediction, based on the assumption that ASR and human intelligibility assessment are related tasks, as both involve speech-to-text mapping \cite{Kolossa22}.

\subsection{Experimental results}
\label{ssec:results}

\begin{table}[t]
	\caption{Performance of the evaluated OIMs in terms of the Pearson correlation coefficient ($\rho$) and Spearman rank correlation coefficient ($r_s$). Best results are marked in bold.}
	\label{tab:overall}
	\centering
	\resizebox{\columnwidth}{!}{
		\begin{tabular}{cccccccc}
			\toprule
			& STOI & ESTOI & STGI & HASPI & SIIB & MOSA-Net+ & W2V-SIP \\
			\midrule
			$\rho$      & 0.94 & 0.89 & \textbf{0.96} & 0.87 & 0.53 & 0.83 & 0.82 \\
			$r_s$ & 0.95 & 0.95 & 0.96 & 0.92 & \textbf{0.97} & 0.84 & 0.93 \\
			\bottomrule
	\end{tabular}}
\end{table}

\begin{table}[t]
	\caption{Performance of the evaluated OIMs per processing condition in terms of the Pearson correlation coefficient ($\rho$) and Spearman rank correlation coefficient ($r_s$). Best results are marked in bold.}
	\label{tab:method}
	\centering
	\begin{tabular}{l c c c c c c}
		\toprule
		& \multicolumn{3}{c}{\emph{Pearson corr. ($\rho$)}} & \multicolumn{3}{c}{\emph{Spearman rank ($r_s$)}} \\
		\cmidrule(lr){2-4} \cmidrule(lr){5-7} 
		Metric & Unp & FSN+ & SGM+ & Unp & FSN+ & SGM+ \\
		\midrule
		STOI      & 0.87 & 0.95 & \textbf{0.99} & \textbf{1.00} & \textbf{0.97} & 0.95 \\
		ESTOI     & 0.82 & 0.93 & \textbf{0.99} & \textbf{1.00} & \textbf{0.97} & \textbf{1.00} \\
		STGI      & \textbf{0.98} & \textbf{0.99} & 0.97 & \textbf{1.00} & \textbf{0.97} & \textbf{1.00} \\
		HASPI     & 0.83 & 0.98 & 0.97 & \textbf{1.00} & \textbf{0.97} & \textbf{1.00} \\
		SIIB      & 0.44 & 0.50 & 0.67 & \textbf{1.00} & \textbf{0.97} & \textbf{1.00} \\
		MOSA-Net+ & 0.83 & 0.94 & 0.97 & 0.85 & 0.96 & 0.86 \\
		W2V-SIP   & 0.87 & 0.67 & 0.92 & 0.99 & 0.94 & \textbf{1.00} \\
		\bottomrule
	\end{tabular}
\end{table}

\begin{table*}[t]
	\caption{Performance of the evaluated OIMs per SNR (dB) in terms of the Pearson correlation coefficient ($\rho$) and Spearman rank correlation coefficient ($r_s$). Note that the coarse set of $r_s$ values results from having only three processing conditions. Best results are marked in bold.}
	\label{tab:snr}
	\centering
	\resizebox{\linewidth}{!}{
		\begin{tabular}{lcccccccccccccccccccccc}
			\toprule
			& \multicolumn{11}{c}{\emph{Pearson correlation coefficient ($\rho$)}} & \multicolumn{11}{c}{\emph{Spearman rank correlation coefficient ($r_s$)}} \\
			\cmidrule(lr){2-12} \cmidrule(lr){13-23}
			Metric & \mbox{-15} & \mbox{-12.5} & \mbox{-10} & \mbox{-7.5} & \mbox{-5} & \mbox{-2.5} & 0 & 2.5 & 5 & 10 & \makebox[1.2em][c]{Clean} 
			& \mbox{-15} & \mbox{-12.5} & \mbox{-10} & \mbox{-7.5} & \mbox{-5} & \mbox{-2.5} & 0 & 2.5 & 5 & 10 & \makebox[1.2em][c]{Clean} \\
			\midrule
			STOI     & 0.34 & 0.62 & 0.95 & \textbf{1.00} & \textbf{0.96} & 0.70 & \mbox{-0.24} & \mbox{-0.37} & 0.17 & \mbox{-0.87} & 0.49
			& 0.00 & \textbf{1.00} & \textbf{1.00} & \textbf{1.00} & \textbf{1.00} & \textbf{1.00} & 0.50 & \mbox{-0.50} & \mbox{-0.50} & \mbox{-1.00} & 0.50 \\
			ESTOI    & 0.98 & 0.98 & 0.98 & 0.96 & 0.78 & 0.24 & \mbox{-0.41} & \mbox{-0.39} & 0.16 & \mbox{-0.86} & \textbf{0.89} & \textbf{0.87} & \textbf{1.00} & \textbf{1.00} & \textbf{1.00} & \textbf{1.00} & 0.50 & 0.50 & \mbox{-0.50} & \mbox{-0.50} & \mbox{-0.50} & \textbf{1.00} \\
			STGI     & 0.94 & 0.96 & \textbf{0.99} & 0.98 & 0.90 & 0.55 & \mbox{-0.32} & \mbox{-0.35} & 0.19 & \mbox{-0.88} & -- & \textbf{0.87} & \textbf{1.00} & \textbf{1.00} & \textbf{1.00} & \textbf{1.00} & 0.50 & 0.50 & 0.50 & \mbox{-0.50} & \mbox{-1.00} & -- \\
			HASPI    & \textbf{0.99} & \textbf{1.00} & 0.85 & 0.78 & 0.52 & \mbox{-0.11} & \mbox{-0.45} & \mbox{-0.35} & 0.20 & \mbox{-0.87} & 0.05 & \textbf{0.87} & \textbf{1.00} & \textbf{1.00} & \textbf{1.00} & 0.50 & 0.50 & 0.50 & 0.50 & \mbox{-0.50} & \mbox{-1.00} & 0.50 \\
			SIIB     & 0.98 & 0.99 & \textbf{0.99} & 0.97 & 0.88 & 0.52 & \mbox{-0.45} & \mbox{-0.48} & 0.11 & \mbox{-0.86} & -- & \textbf{0.87} & \textbf{1.00} & \textbf{1.00} & \textbf{1.00} & \textbf{1.00} & 0.50 & 0.50 & \mbox{-0.50} & \mbox{-0.50} & \mbox{-0.50} & -- \\
			MOSA-Net+ & \mbox{-0.97} & \mbox{-1.00} & \mbox{-0.96} & \mbox{-0.93} & \mbox{-0.44} & \textbf{0.94} & \textbf{0.92} & \textbf{0.65} & \textbf{0.48} & \mbox{-0.80} & 0.37 & \mbox{-0.87} & \mbox{-1.00} & \mbox{-1.00} & \mbox{-0.50} & \mbox{-0.50} & \textbf{1.00} & \textbf{1.00} & \textbf{1.00} & \textbf{0.50} & \mbox{-0.50} & 0.50 \\
			W2V-SIP  & 0.69 & 0.74 & 0.94 & 0.98 & \mbox{-0.97} & \mbox{-0.31} & 0.03 & \mbox{-0.16} & \mbox{-0.64} & \textbf{0.47} & 0.82 & \textbf{0.87} & \textbf{1.00} & \textbf{1.00} & \textbf{1.00} & \mbox{-1.00} & \mbox{-0.50} & \mbox{-0.50} & \mbox{-0.50} & \mbox{-0.50} & \textbf{0.50} & 0.50 \\
			\bottomrule
	\end{tabular}}
\end{table*}

For OIMs, a key requirement is a strong positive monotonic relationship with behavioral intelligibility. Accordingly, performance is typically assessed using the Pearson correlation coefficient ($\rho$) and the Spearman rank correlation coefficient ($r_s$), which we also adopt in this work\footnote{Specifically, Spearman correlations were computed using tied ranks.}. While $\rho$ measures linear correlation between objective and human scores, $r_s$ captures their monotonic relationship. Because the mapping between raw OIM scores and behavioral intelligibility is often nonlinear, prior work has applied logistic mapping before computing $\rho$ \cite{VanKuyk}. However, following \cite{Haolan26}, we do not apply such mapping, as behavioral intelligibility scores are unavailable in practical scenarios. Consequently, $r_s$ is treated as the primary evaluation metric.

Table \ref{tab:overall} reports the performance of the evaluated OIMs in terms of $\rho$ and $r_s$. Figure \ref{fig:scatter} shows scatter plots of behavioral intelligibility versus OIM scores, where each data point corresponds to a combination of SNR and processing condition. Metric-wise logistic fits are also included. As shown in Table \ref{tab:overall}, SIIB achieves the best performance on SpInt in terms of the Spearman rank correlation coefficient (0.97), despite exhibiting the lowest \emph{linear} correlation with behavioral intelligibility. The latter is primarily due to the clean speech condition ($+\infty$ dB input SNR), which yields SIIB values above 1,200 b/s (see Figure \ref{fig:scatter}). The strong performance of SIIB in terms of $r_s$ is consistent with \cite{SIIB}, where the authors report a high correlation between SIIB and the intelligibility of speech degraded by additive noise and processed by enhancement algorithms. SIIB was developed and evaluated using multiple speech intelligibility datasets covering Dutch, Danish, and English speech \cite{VanKuyk}. Furthermore, STGI and STOI---originally developed for evaluating noise-reduction algorithms---exhibit strong performance in terms of both $\rho$ and $r_s$.

Regarding the two no-reference OIMs assessed, Table \ref{tab:overall} and Figure \ref{fig:scatter} show that W2V-SIP outperforms MOSA-Net+. Although MOSA-Net+ leverages a powerful multilingual ASR model such as Whisper for feature extraction---trained on data that include Spanish---it was fine-tuned for speech intelligibility prediction using Taiwanese Mandarin data \cite{MOSA-Net+}. In contrast, W2V-SIP was developed using English data only \cite{Haolan}. Given that Spanish is linguistically closer to English than to Taiwanese Mandarin, this training--test mismatch may partly explain the poorer performance of MOSA-Net+.

Table \ref{tab:overall} reveals that intrusive OIMs tend to outperform their deep learning-based no-reference counterparts. We hypothesize that this is due to the former's access to clean reference signals, which makes reference-based OIMs more robust to training--test acoustic mismatches, such as language mismatch \cite{Pedersen23,Haolan,Haolan26}. This is particularly relevant in our scenario, as, to the best of our knowledge, none of the evaluated OIMs were developed using Spanish data.

Additionally, Table \ref{tab:method} reports the performance of the evaluated OIMs per processing condition---unprocessed (Unp), FullSubNet+ (FSN+), and SGMSE+ (SGM+)---in terms of both $\rho$ and $r_s$. Per-SNR performance is similarly reported in Table \ref{tab:snr}. Taken together, these results indicate that OIMs generally capture the impact of SNR on intelligibility (particularly the intrusive ones; see Table \ref{tab:method}), but struggle to reliably rank processing conditions, as reflected by the small and, in several cases, strongly negative $\rho$ and $r_s$ values in Table \ref{tab:snr}. Indeed, there are cases where metrics yield a perfect inverse ranking ($r_s=-1.00$), i.e., opposite to human intelligibility judgments (e.g., MOSA-Net+ at -12.5 and -10 dB, or STOI, STGI, and HASPI at 10 dB). This outcome is in agreement with previous work \cite{Espejo23}, where STOI, ESTOI, and STGI were found to exhibit a negative monotonic relationship with behavioral intelligibility for certain combinations of noise type and SNR when evaluated over pooled data from multiple deep learning-based speech enhancement systems \cite{Kolbaek20,Espejo23}. The authors of \cite{Espejo23} therefore caution against relying solely on OIMs for comparing speech enhancement systems, noting that human intelligibility assessments cannot be systematically replaced by objective metrics.

\section{Conclusion}
\label{sec:conclusion}

In this study, we have evaluated the performance of five intrusive (STOI, ESTOI, STGI, HASPI, and SIIB) and two non-intrusive (MOSA-Net+ and W2V-SIP) OIMs on a new Spanish speech intelligibility dataset, SpInt. The results have shown that intrusive OIMs tend to outperform their deep learning-based non-intrusive counterparts. We have attributed this to the former's access to clean reference signals, which makes them more robust to training--test acoustic mismatches, such as language mismatch---particularly relevant in our experimental setting. As a result, to support the development of more robust and generalizable deep learning-based no-reference OIMs, we have made SpInt publicly available.

\section{Acknowledgments}
This work was supported by the Spanish Ministry of Science and Innovation under the ``Ram\'on y Cajal'' programme (RYC2022-036755-I). We also thank H\"{o}rzentrum Oldenburg gGmbH for providing the Spanish speech material used in the intelligibility test, and we sincerely appreciate all participants who contributed by taking the test.

\bibliographystyle{IEEEtran}
\bibliography{mybib}

\end{document}